# Graphene as Infrared Light Sensor Material

Ahalapitiya H. Jayatissa[a)] and Madhav Gautam
Mechanical, Industrial, and Manufacturing Engineering (MIME) Department
The University of Toledo, OH 43606, USA
[a)]Correspondence: ahalapitiya.jayatissa@utoledo.edu

**Abstract**: The infrared (IR) photoresponse of graphene synthesized by atmospheric chemical vapor deposition (CVD) system using a mixture of hydrogen and methane gases was studied. The IR sensor devices were fabricated using graphene films transferred on to a $SiO_2$ substrate by a lift off process. The quality of graphene was investigated with the Raman spectroscopy and optical microscopy. The photoresponse was recorded under the illumination of IR light of wavelength 850 nm and intensity of around 2.16 µW/mm$^2$. The effects of temperature and hydrogenation on photoconductivity were also studied. It was found that the transient response and recovery times decreased with the increase of the temperature. Hydrogenation effect also caused the significant decrease in the photoresponse of the device. Although the net change in the photoresponse for IR light was lower at low illumination intensity levels, the transient responses were observed around 100 times faster than the recently reported CNT-based IR sensors.

## 1. Introduction

Optoelectronic devices working in near infra-red (NIR) (800 - 2000 nm) are always demanding for different applications [1-4]. There has been significant works reported on the fabrication of optoelectronic devices using NIR materials [5-12]. In recent years, single walled carbon nanotubes (SWCNTs) have been investigated extensively as a semiconducting material for IR sensors because of its strong absorption behavior in NIR region [7-12]. One of the key challenges in developing NIR detectors is the finding of ultra fast optical response in the sensor material [5-8]. Recently, strong absorption behavior in NIR region has been reported for thermally reduced graphene oxides [1,2]. This provides a pathway to use graphene as an optoelectronic material for IR detection. Although the optical properties of graphene in visible region have been reported by many researchers [13-15], we have not found any research work related to the photoresponse of graphene in IR region of the spectrum. In this paper, photoresponse of graphene film on macro-scale has been reported in different conditions.

Graphene is a monolayered carbon film with a film thickness of around 0.32Å [13 - 15], where carbon atoms are arranged in a two-dimensional hexagonal lattice structure. It can be thought of as a single layer peeled off from the graphite stack. It has evolved as an interesting material due to its unique physical and electrical properties [16]. This material is different from most of the conventional semiconductors because of its zero bandgap semi-conducting behavior [17]. For example, graphene-based transistor devices may operate very faster than traditional silicon devices due to high intrinsic carrier mobility (~ $2 \times 10^5$ cm$^2$v$^{-1}$s$^{-1}$) [1, 2, 18]. Being the material of high mechanical stress and low density (2.2 gm/cm$^3$), it may lead to the application in nano-robotics [19, 20].

We have investigated the photoconductivity of graphene layers synthesized in atmospheric chemical vapor deposition (CVD) of $CH_4$ on a copper substrate. The devices were fabricated by transferred CVD graphene onto a $SiO_2$/Si substrate. The investigations were carried out to understand the temperature dependence and hydrogenation effect on photoconductivity of graphene in NIR region. Although the net change in the photoresponse for IR light was lower at low illumination intensity levels (2.16 µW/mm$^2$), the transient responses were observed around 100 times faster than photoconductivity of CNT for NIR lights.

## 2. Experimental Procedures

The growth of graphene films was carried out on a copper (Cu) substrate (25 µm thick) in an alumina tube furnace system under the flow of methane ($CH_4$) and hydrogen ($H_2$) gases. Copper substrate (99.999% pure, Alfa Aesar) was heated in a tube furnace under the 150 standard cubic centimeters per minute (sccm) flow of mixture of hydrogen and Argon (10% $H_2$, 90% Ar) and annealed at 1100 $^0$C for one hour. After annealing, graphene deposition was carried out by passing a mixture of methane and argon (5% $CH_4$, 95% Ar) followed by the immediate cooling. Graphene deposited on copper by CVD method was transferred to $SiO_2$/Si substrate by wet etching of Cu [15, 21-23]. The thickness of the thermally-grown $SiO_2$ was 118 nm as confirmed by UV spectrometry [24]. The Raman spectra of these films were recorded with the excitation wavelength of 530 nm.

In order to fabricate the IR sensors, a thin layer of gold (about 100 nm) was coated onto the transferred graphene film by a vacuum evaporation method. The gold electrodes were patterned by lithography followed by etching of gold with aqueous $KI/I_2$ solution. The spacing and the length of these electrodes were 6 mm and 4 mm, respectively. Fig. 1 shows the schematic diagram of the fabricated IR sensor and photoresponse measurement circuit. The device was biased with a constant voltage (1.0 V) during collection of the data. To understand the reflection of light from graphene, reflectance from bi-layer substrate ($SiO_2$/Si) and tri-layer substrate (graphene/$SiO_2$/Si) were measured with a double beam UV/Visible spectrometer (Shimadzu). The reflectance spectra were investigated in the spectral range 300-1100 nm.

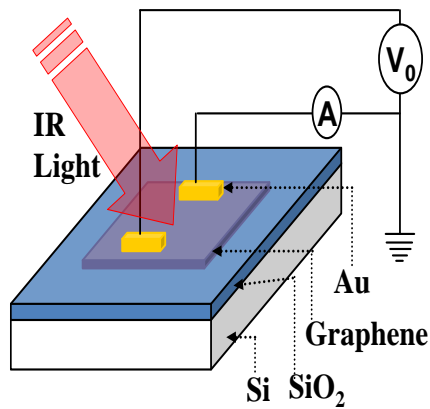

Fig.1: Schematic of photoresponse measurement system ($V_0$= 1.0 V).

## 3. Results and Discussions

### 3.1. Surface Characterization

The Raman spectroscopy has been used to characterize the quality of graphene. The Raman spectrum of Graphene gives for main bands corresponding to the vibration mode of graphene. Fig. 2 shows the as-measured Raman spectra of graphene films produced on $SiO_2$ surface. The spectrum was normalized with respect to the intensity level of 2D band. The peak at around 1580 $cm^{-1}$ and 2660 $cm^{-1}$, respectively, indicate the G band and the 2D band, which are characteristics Raman peaks of graphene. It has been reported that the defect free monolayer graphene can be identified with characteristic features of Raman band intensities [25]. The intensity of 2D band is ~2 times larger than the intensity of G band suggesting that the presence of less defective graphene on $SiO_2$ surface. This fact is also supported by the weak intensity of D-band (1350 $cm^{-1}$).

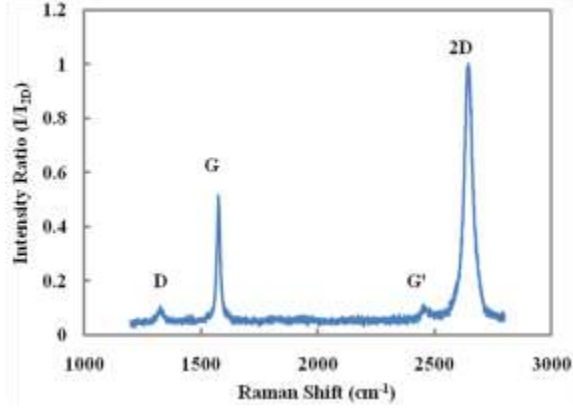

**Fig. 2: Raman spectra of graphene transferred to silicon wafer (SiO$_2$ + Si) scaled with respect to the maximum peak.**

*3.2. Photoconductivity*

*3.2.1. Dynamic response*

Fig. 3 shows the dynamic response of photoconductivity of graphene film for the NIR light at room temperature. Fig. 3(a) shows the response and recovery of the device when the IR light was turned on and off, respectively, whereas Fig. 3(b) indicates the same characteristic for one cycle only. The intensity of the IR light source used was 2.16 µW/mm$^2$ at the device surface. Although the intensity level was very low, a clear photoresponse of device was measured. The photogeneration of carriers can be primarily attributed to the creation of bands at the defect of graphene sheets. When graphene is deposited on a copper plate, defects are developed at the grain boundary of polycrystalline copper films. We believe that these defects are responsible for the creation of localized photoactive regions, which contribute to the photogeneration of carriers [26,27]. The photoresponse could be characterized with a time step function. In both the photocurrent increase and drop cases, the experimental data were fitted well into the exponential form as [10],

$$I = I_o + A_o \exp\left(\frac{-t}{\tau}\right). \tag{1}$$

Here, *I* is the current, *t* is the response time and $I_o$, $\tau$ and $A_0$ are constants. Fig. 4(a) and 4(b) show the fit of the response in the form explained above. The data analysis indicated that the time constants were 10 ms and 31 ms for rise and fall of the photocurrent, respectively.

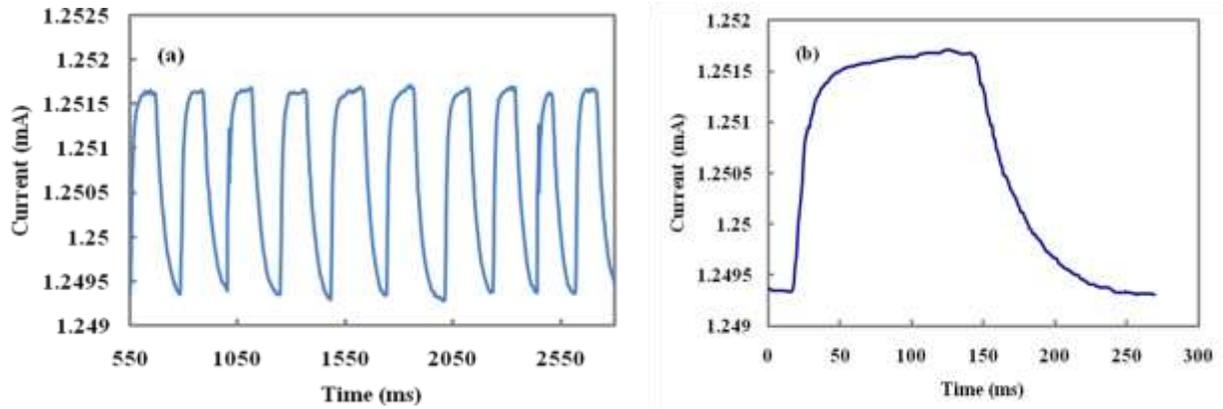

**Fig. 3: The photoresponse of the device due to IR light for (a) different cycles and (b) for one cycle.**

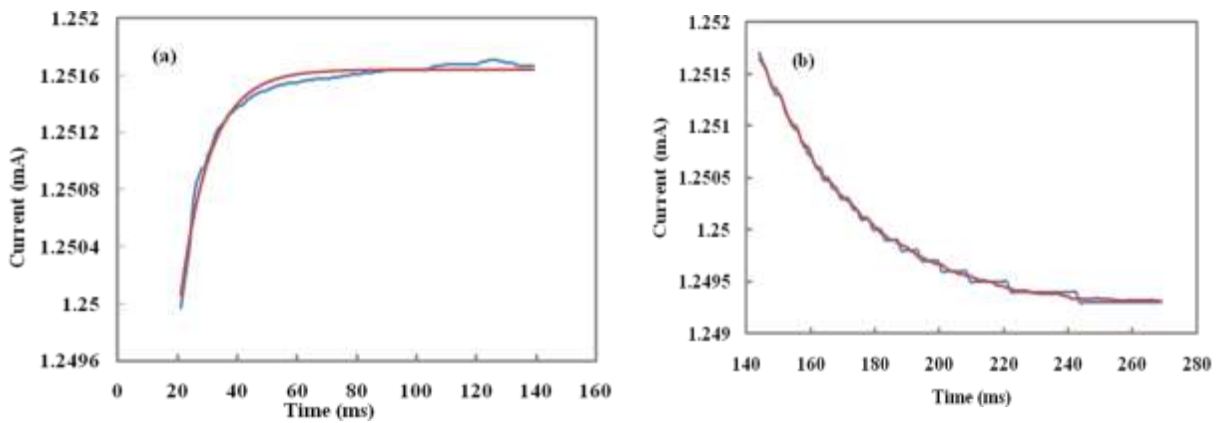

**Fig. 4: The photoresponse of the device due to IR light for (a) response and (b) recovery.**

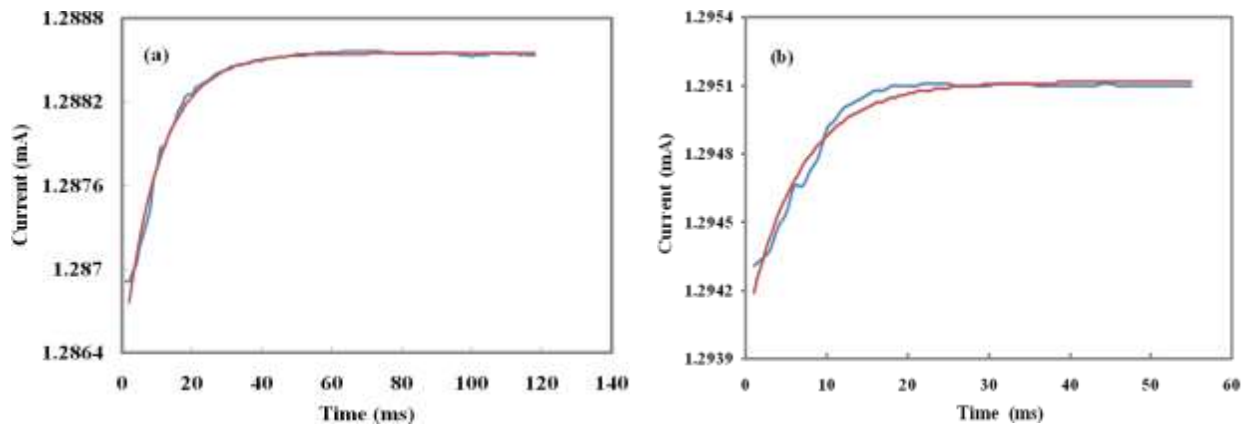

**Fig. 5: The photoresponse of the device due to IR light at (a) 50 °C and (b) 100 °C.**

### 3.2.2. *The effect of temperature on photoconductivity*

Fig. 6 shows the effect of temperature on the photoconductivity of graphene. The photoconductivity was tested at 50 °C and 100 °C, respectively. During the experiment, the device was heated to the desired temperature for 30 minutes to ensure the thermal equilibrium. Transient responses of the device were 10.26 ms and 6.57 ms and the transient recovery times were 12.55 ms and 5.91 ms at 50 $^0$C and 100 $^0$C, respectively. A significant difference in transient response of the device was not found when the device temperature was increased from room temperature to 50 $^0$C and transient response time decreased by 40% when the temperature was changed from 50 $^0$C to 100 $^0$C. Similarly, the transient recovery time decreased by 60% when the temperature was changed from room temperature to 50 $^0$C and it decreased by 50% when the temperature was changed from 50 $^0$C to 100 $^0$C.

On the other hand, the amplitude of the photocurrent didn't show any significant difference when the temperature was changed from room temperature to 50 $^0$C whereas it decreased by 50% when the temperature was changed from 50 $^0$C to 100 $^0$C. A slight change in photocurrent at high temperature measurement (100 $^0$C) from low temperature (50 $^0$C) can be attributed to the career generation is influenced by thermal effect associated with defects. Furthermore, the increase in current due to the thermal effect of IR light is less pronounced at elevated temperatures because the change in the temperature by IR heating is negligible. Therefore, the total photocurrent generation can be attributed to the photo generation of carriers in the graphene.

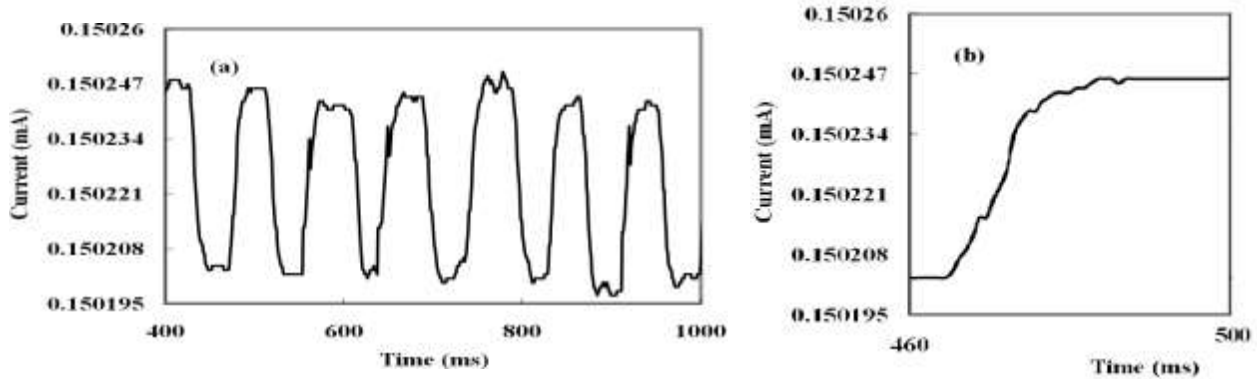

**Fig. 6: The photoresponse of the device in IR light due to hydrogenation at 100sccm of hydrogen flow for (a) difference cycle and (b) one cycle.**

On the other hand, the amplitude of the photocurrent didn't show any significant difference when the temperature was changed from room temperature to 50 $^0$C whereas it decreased by 50% when the temperature was changed from 50 $^0$C to 100 $^0$C. Smaller change in low temperature gradient can be attributed to the fact that small bandgap in graphene. Furthermore, the increase in current due to the thermal effect of IR light is less pronounced at elevated temperatures because the change in the temperature by IR heating is negligible. Therefore, the total photocurrent generation can be attributed to the photo generation of carriers in the graphene.

*3.2.3. The effect of hydrogenation on photoconductivity*

The effect of hydrogenation on photoresponse of the device was tested at 100 $^0$C for different concentrations of hydrogen flow rates. The device was heated at 100 $^0$C for 30 min to ensure the thermal equilibrium followed by the constant hydrogen flow for more than one hour until reach of the saturation of surface of graphene by hydrogen by adsorption. The saturation was confirmed by monitoring resistance changes against time using two-point probe method.

Fig. 7 shows the photoresponse of the device at different flow rates of hydrogen. Transient responses of the device were 6.05 ms and 7.27 ms in 50 sccm and 100 sccm flow rate of hydrogen gas, respectively, and the corresponding values during recovery process were 7.1 ms and 7.81 ms, respectively. The transient response of the device was found to differ by 17% in going from 50 to 100 sccm of hydrogen flow rates. Table 1 lists the transient response and recovery times at different temperatures to compare the effect of hydrogenation.

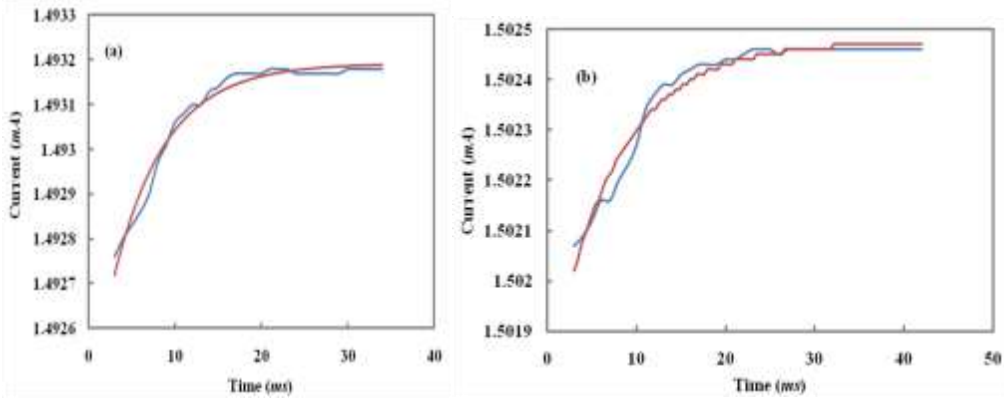

**Fig. 7: The photoresponse of the device in IR light due to hydrogenation at (a) 50 sccm and (b) 100 sccm flow rate of hydrogen gas at 100 $^0$C.**

**Table 1: Transient response and recovery times at different temperatures**.

| Temperature ($^0$C) | Transient response ($\tau_1$) (ms) | | Transient recovery ($\tau_2$) (ms) | |
|---|---|---|---|---|
| | In vacuum | In hydrogen (100 sccm) | In vacuum | In hydrogen (100 sccm) |
| Room Tem. | 10.04 | 13.90 | 31.26 | 44.29 |
| 100 | 6.57 | 7.24 | 5.91 | 7.81 |

The photoresponse of the device in hydrogen was also calculated and compared with that of the device in vacuum at different temperatures. Response of the device was calculated using the formula given by [25],

$$S = \left(\frac{I_1 - I_2}{I_2}\right) * 100\% . \qquad (2)$$

Where, $I_1$ and $I_2$ are the currents with and without IR light, respectively. Generally, response is calculated in percentage.

Fig. 8 shows the comparison of the responses due to hydrogenation effect at 100 $^0$C. The response was found to decrease by around 57% when the device was hydrogenated at 50 sccm flow rate of hydrogen gas while it decreased by around 68% when the flow rate was increased to 100 sccm. The effect of hydrogenation was even seen substantial at room temperature compared with hydrogenation at 100 $^0$C. The flow of hydrogen was continued during cooling. The decrease in the response of the device due to hydrogenation effect was observed as expected. The semiconducting Behaviour of graphene is attributed to the formation of bands at the defect sites [26]. When hydronation is occurred, the conductivity can be reduced to a certain extent due to the passivation of defect sites with hydrogen.

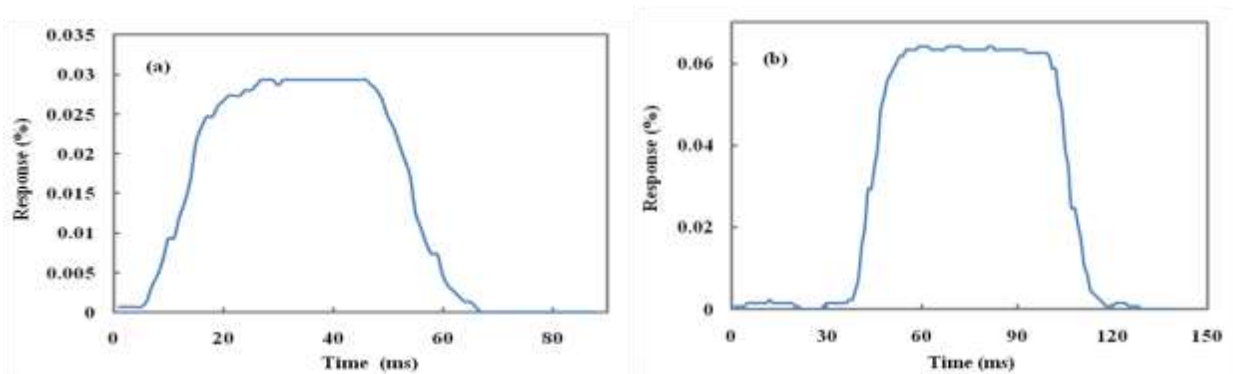

**Fig. 8: The photoresponse of the device in IR light at 100 $^0$C in (a) hydrogenation at 100 sccm of hydrogen flow and (b) in ambient condition.**

## 4. Conclusion

In this paper, a graphene-based IR sensor was investigated in different conditions in terms of the photoresponse in the presence of light. The device was fabricated between electrode materials and the presence of a monolayer of graphene was confirmed by Raman Spectroscopy. The effect of temperature on photoconductivity was recorded at different temperature conditions. The photoconductivity of graphene films was interpreted as due to the creation of localized bands in defect sites at the gran boundaries of CVD graphene. The device exhibited a temperature-dependent effect on the photoresponse behavior. The transient response and recovery times were seen reduced in the high-temperature region, indicating that the thermal effect due to heating was more pronounced than the heating effect caused by the IR light. It also revealed the fact that the net photocurrent change due to IR light decreases as the charge carriers responsible for conduction are already excited to the conduction band due to thermal heating before IR light was used. The hydrogenation effect on photoconductivity was also studied. The hydrogenation caused a significant decrease in the photoresponse of the device at high temperature as expected because the hydrogen ions were believed to be adsorbed at the grain boundaries and passivate the defects that are responsible for photoconductivity. As the device was illuminated with a low intensity (~ 2.16 µW/mm$^2$) of IR light, the net change in the photocurrent was not significant. However, the transient responses were observed around 100 times faster than the recently reported CNT-based IR sensor, which may lead to the application of graphene towards ultra-fast optical response devices.

**Acknowledgements:** This research was supported by a grant (Grant #: ECCS 0925783) from National Science Foundation (NSF) of USA.